# Vertical stratified turbulent transport mechanism indicated by remote sensing

Carl H. Gibson, R. Norris Keeler, Valery G. Bondur

*Satellite and shipboard data reveal the intermittent vertical information transport mechanism of turbulence and internal waves that mixes the ocean, atmosphere, planets and stars.*

Astronauts noticed they could see deep bottom features and internal waves within the ocean from space platforms (Figure 1), but were not believed because no known physical mechanism permits remote sensing through kilometers of opaque water. The information transport mechanism is generic to all large stratified intermittently turbulent bodies of natural fluid[1–8]. Vertical (radial) heat, mass, momentum and energy transport are affected. The mechanism involves turbulence[1], fossil turbulence[2] and nonlinear internal waves[3]. A three year series of oceanographic experiments were organized to test Russian claims they could detect submerged turbulence, internal waves, bottom depth and topography from space satellites.

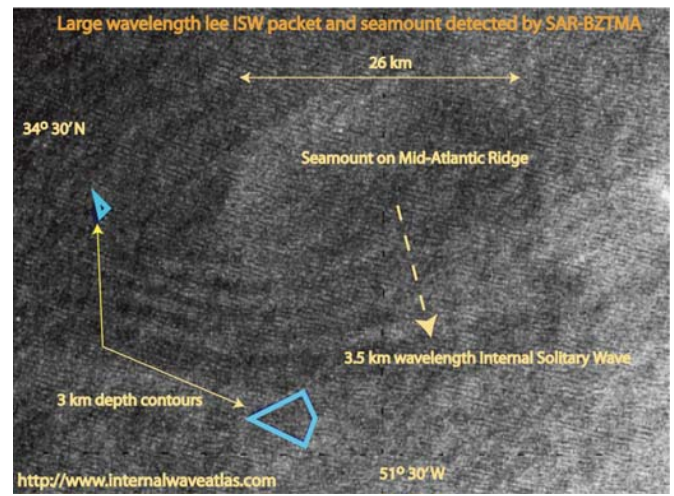

Figure 1. Seamount and internal tidal waves from space. How is this information transmitted?

The Remote Anthropogenic Sensing Program (RASP 2002, 2003, 2004) used the turbulent Honolulu wastewater outfall at Sand Island with full sea truth during space satellite observations. Buoyancy traps wastewater at depth 50 meters to prevent contamination of surface waters and beaches. Remarkably, outfall fossil and zombie turbulence[4] patches can dominate mixing in Mamala Bay at distances exceeding 20 km in areas exceeding 200 $km^2$.

**The mechanism**

---

[1] Turbulence is defined as an eddy-like state of fluid motion where the inertial-vortex forces $\vec{v} \times \vec{\omega}$ of the eddies are larger than any other forces that tend to damp the eddies out, where $\vec{v}$ is the velocity and $\vec{\omega}$ is the vorticity. Thus, turbulence always cascades from small scales to large.

[2] Fossil turbulence is defined as a perturbation in any hydrophysical field produced by turbulence that persists after the fluid ceases to be turbulent at the scale of the perturbation.

[3] When active turbulence cascades to the (Ozmidov scale) maximum vertical size permitted by buoyancy $L_R = (\varepsilon / N^3)^{1/2}$ it fossilizes, and the turbulent kinetic energy is radiated near vertically as fossil and zombie turbulence waves.

[4] Zombie turbulence is produced when density gradients of fossil turbulence patches are tilted by internal waves to create vorticity at rates $(\nabla \rho \times \nabla p) / \rho^2$.

Figure 2 shows RASP data and how it was obtained and analyzed. Brightness anomaly wavelengths suggest Ozmidov scales at fossilization from strong turbulence events radiate narrow spatial-frequency packets of fossil turbulence waves near-vertically. Detected wavelengths (40-160 meters) are confirmed by thermistor-chain internal-wave measurements and surface-wave detectors. Horizontally towed and vertically dropped microstructure profilers contoured viscous and temperature dissipation rates and tested for hydrodynamic states of the patches. Intermittent mixing chimneys were detected.

The mechanism for vertical transport is termed "Beamed Zombie Turbulence Maser Action" (BZTMA), and occurs in intermittent mixing chimneys. The outfall fossil turbulence patches drift away from their source with ambient currents, and absorb energy from bottom generated fossil turbulence waves to form zombie turbulence and zombie turbulence waves in an efficient maser action that moves length-scale-information for the bottom waves to the sea surface where the zombie-turbulence-waves (ZTWS) break and permit its detection. Vertical ZTW chimneys follow paths of previous ZTWs, just as lightning flashes follow ionized paths of previous lightning flashes. The BZTMA mechanism is an extension of Kolmogorovian universal similarity theory with extensions to scalar mixing [2]. Consequently it is generic to natural fluids such as the ocean, atmosphere, planets and stars.

**Astrophysical implications**

Application of modern fluid mechanics to astrophysics [8] reveals the dark matter of galaxies as thirty million Primordial-Fog-Particle earth-mass frozen-gas planets (PFPs) per star in Proto-Globular-star-Cluster-mass clumps (PGCs), but requires the use of BZTMA to explain why stars form and die in different ways and why it is not necessary to believe in dark energy. Figure 3 shows 12 decades of Kolmogorov-Corrsin-Obukhov electron density spectra from earth scales to PGC scales that require planetary atmospheres evaporated by nearby supernovae (II) and their resulting pulsars. This requires strong BZTMA mixing by planet accretion. Otherwise the carbon cores of stars will not be mixed away, giving supernovae Ia.

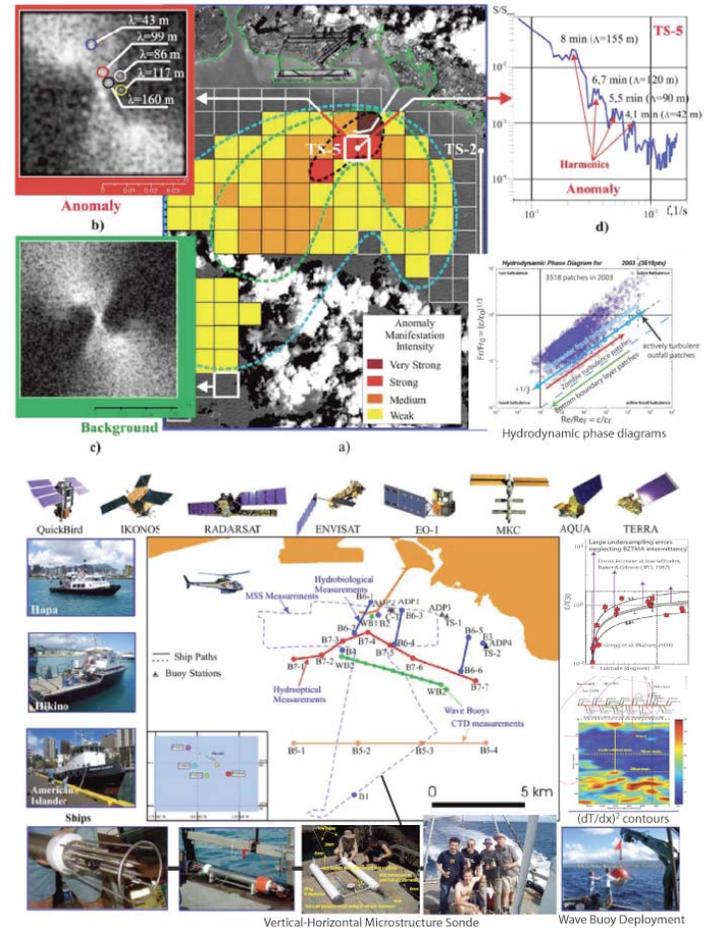

Figure 2. Sea surface brightness anomalies are detected [7] (top) by comparing 2D-spectra near the outfall (red) with background regions (green). Corresponding internal wavelengths (right) suggest fossil turbulence waves radiated from the bottom. Key RASP microstructure results are shown (bottom), with ships, satellites, microstructure detectors, platforms, and section paths.

**Conclusions**

Stratified turbulent mixing in natural fluids is dominated by the most powerful turbulent events of a vertical column, which fossilize and radiate nonlinear internal waves near-vertically. These mix to form fossil turbulence patches, secondary zombie turbulence events, mixing chimneys and information transport to the surface by the BZTMA mechanism. This explains why deeply submerged seamounts and internal waves (Fig. 1) can be seen and why vast undersampling errors are typical in physical oceanography when fossil turbulence and fossil turbulence waves are neglected.

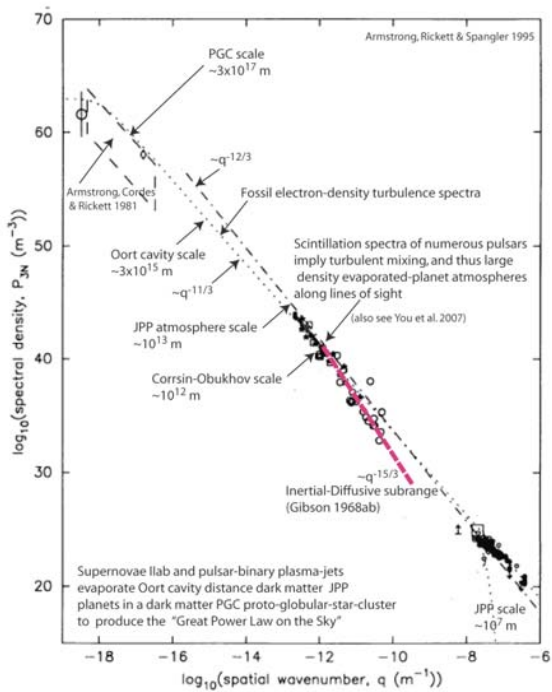

Figure 3. Application of BZTMA mixing theory to understand pulsar electron density fluctuation spectra and star formation from planets. Jovian PFP (primordial–fog-particle) Planets (JPPs) comprise the baryonic dark matter of all galaxies and develop turbulent atmospheres when evaporated by radiation from rapidly spinning white dwarf and neutron stars.

## Author information

Prof. Carl H. Gibson is a faculty member of MAE and SIO Departments of UCSD and a member of the Center for Astrophysics and Space Sciences. He specializes in turbulence, fossil turbulence, turbulent mixing, and turbulent transport processes in the ocean, atmosphere, planets, stars, and cosmology. With Norris Keeler and Valery Bondur, he proposed tests, using the Honolulu municipal outfall, of Russian claims that submerged turbulence can be detected remotely from space satellites. The successful RASP 2002, 2003 and 2004 expeditions are reported in the SPIE 2007 CORS 6680 conference. Scripps Institution of Oceanog. Dept.,University of Cal. San Diego, La Jolla CA 92093-0411
USA, 858 534-3184. cgibson@ucsd.edu, sdcc3.ucsd.edu/~ir118

Dr. R. Norris Keeler headed the Physics Department of Lawrence Livermore National Laboratory, was President of the International High Pressure Society, Director of Materiel for the US Navy, and is now a Director of Directed Technologies, Inc. He initiated tests of Russian claims they could detect submerged turbulence by remote sensing, resulting in the RASP 2002, 2003 and 2004 expeditions and many scientific papers.
norris_keeler@directedtechnologies.com

Prof. Valery G. Bondur directs a large group of Russian specialists in remote sensing of geophysical parameters. He is a member of the Russian Academy of Sciences and President of the International Eurasian Academy of Sciences. He helped design and carry out the international Remote Anthropogenic Sensing Program (RASP) 2002, 2003 and 2004 expeditions reported in SPIE Coastal Ocean Remote Sensing Proceedings 6680, Aug. 2007. Aerocosmos Scientific Ctr. of Aerospace Monitoring, Moscow, Russia, vgbondur@online.ru

We wish to acknowledge Robert Arnone's invitation (to RNK) to contribute [1] to CORS 6680.